\documentclass[letterpaper]{sig-alternate}
\usepackage{graphicx}
\usepackage{graphics}
\usepackage{wrapfig}
\usepackage{algorithmic}
\usepackage{algorithm}
\usepackage[all]{xy}
\long\def\comment#1{}
\long\def\remove#1{}
\newcommand{\commentout}[1]{}
\newcommand{\denselist}{
      \setlength{\itemsep}{0pt}
      \setlength{\parsep}{1.5pt}
      \setlength{\topsep}{1.5pt}
      \setlength{\parskip}{2pt}
      \setlength{\partopsep}{0pt}
      \setlength{\labelwidth}{1em}
      \setlength{\labelsep}{0.5em} }
\newcommand{\bdesc}{\begin{itemize}\denselist}
\newcommand{\edesc}{\end{itemize}}

\let\oldbibliography\thebibliography
\renewcommand{\thebibliography}[1]{%
  \oldbibliography{#1}%
  \setlength{\itemsep}{-2pt}%
}

\conferenceinfo{WSDM'11,} {February 9--12, 2011, Hong Kong, China.}
\CopyrightYear{2011}
\crdata{978-1-4503-0493-1/11/02}
\begin{document}
\title{A Framework for Quantitative Analysis of Cascades on Networks}

\numberofauthors{2}
\author{
\alignauthor Rumi Ghosh\\
       \affaddr{USC Information Sciences Institute}\\
       \affaddr{4676 Admiralty Way}\\
       \affaddr{Marina del Rey, CA 90292}\\
       \email{rumig@usc.edu}
\alignauthor
Kristina Lerman\\
       \affaddr{USC Information Sciences Institute}\\
       \affaddr{4676 Admiralty Way}\\
       \affaddr{Marina del Rey, CA 90292}\\
       \email{lerman@isi.edu}
}

\maketitle

\begin{abstract}
How does information flow in online social networks? How does the structure and size of the information cascade evolve in time? How can we efficiently mine the information contained in cascade dynamics? We approach these questions empirically and present an efficient and scalable mathematical framework for quantitative analysis of cascades on networks. We define a \emph{cascade generating function} that captures the details of the microscopic dynamics of the cascades. We show that this function can also be used to compute the macroscopic properties of cascades, such as their size, spread, diameter, number of paths, and average path length. We present an algorithm to efficiently compute cascade generating function and demonstrate that while significantly compressing information within a cascade, it nevertheless allows us to accurately reconstruct its structure. We use this framework to study information dynamics on the social network of Digg. Digg allows users to post and vote on stories, and easily see the stories that friends have voted on. As a story spreads on Digg through voting, it generates cascades. We extract cascades of more than 3,500 Digg stories and calculate their macroscopic and microscopic properties. We identify several  trends in cascade dynamics: spreading via chaining, branching and community. We discuss how these affect the spread of the story through the Digg social network. Our computational framework is general and offers a practical solution to quantitative analysis of the microscopic structure of even very large cascades.
\end{abstract}

\keywords{information spread, cascades, diffusion, online social networks}

\section{Introduction}
Throughout history, the flow of  ideas and innovation has led to vast cultural, economic, and political changes. Social scientists have studied  this phenomenon in detail in several different settings~\cite{Rogers} and found that ideas and innovations tend to diffuse along social links. First, an innovator adopts a novel idea or practice, then people in contact with the innovator adopt it, then people in contact with those people, and so on. In this way, information cascades on a social network. Not surprisingly, information cascades are also common in online social networks.  They are created, for example, when an individual forwards an email she receives to her contacts~\cite{Wu03,Kleinberg-memes}, or retweets a news item to her followers on Twitter~\cite{Lerman10icwsm}. Understanding how information spreads in online social networks may be indicative of its quality~\cite{Crane08SIPS,Lerman08wosn}. A mathematical tool for analysis of cascades can find extensive use anywhere where cascades are studied: anomaly and spam detection, information classification, viral marketing, epidemiological studies, computer virus spread, political and social unrest and even power transmission failure~\cite{Dodds04, Watts02}.

Availability of large scale data about human behavior in online social networks has enabled computational scientists to investigate what drives information diffusion and suggest mechanisms to facilitate its spread. However, as in any other field of research, there are two distinct ways of tackling this problem:  model-centric or empirical. Model-centric approaches make certain assumptions about how individuals participating in a cascade are affected by their neighbors (independent cascade or threshold model). Using these models, researchers have tried to infer global properties of information cascades in social networks~\cite{Young02,Watts02}, devise efficient methods to infer the underlying network structure~\cite{Rodriguez10,Gruhl04} or maximize cascade size~\cite{Richardson01,Kempe03}, and identify influential spreaders~\cite{Kitsak10}. However, empirical approaches are needed to validate assumptions made by these models. We need principled mathematical tools to quantitatively characterize the temporal and spatial properties of cascades as they occur in real-world networks. However, to the best of our knowledge, no previous work has attempted to quantify the dynamics of information cascades on social networks or characterize their microscopic growth. At most, researchers have visualized the shape of cascades~\cite{Kleinberg-memes} or enumerated their commonly observed patterns~\cite{Leskovec07}. Such approaches do not scale to even moderately large cascades.

To address this gap, we propose a \emph{practical, general, and scalable quantitative framework} for the analysis of cascades on social networks that is applicable even to large cascades. We define a \emph{cascade generating function}, which captures the details of the dynamics of information diffusion on networks. We can use this function to (1) compute the macroscopic properties of the cascade, such as its size, diameter, average path length, etc., (2)  reconstruct the shape of the cascade, and (3) analyze its microscopic dynamic properties. The cascade generating function is a good \emph{signature}~\cite{Leskovec05} of the contagion process occurring on a network.  It could help us identify patterns, trends, and anomalies within the cascades in near {real-time}. It could aid spam filtering, since the flow of spam messages within a network will be different from the flow of valid information. It could be useful for viral marketing, since it can help us discover the signature of trends that become popular as compared to those which do not.

As the size of cascades grows,  storing their complete structure  may not be feasible. However, the cascade generating function can approximate the structure of the cascade with very high accuracy, in spite of having  pseudo-linear space complexity. Hence, the cascade generating function can provide \emph{efficient compression} of the information in a cascade.

This paper makes the following contributions. In Section~\ref{sec:framework} we describe a general mathematical framework for representing and quantitatively analyzing cascades on social networks. Specifically, in Section~\ref{sec:characterization}, we define the cascade generating function, which describes how information spreads through the network. We show that this function can be used to compute cascades' macroscopic properties, such as its size, diameter, number of paths in the cascade, etc. In Section ~\ref{sec:comp}, we present a fast, efficient algorithm to compute this function, having $O(kdN)$ runtime complexity and $O(kN)$ space complexity in its naive implementation, where $N$ is the number of nodes participating in a cascade, $d$ is the maximum degree of any node and $k$ is the number of independent cascade seeds.  We demonstrate the use of cascade generating function to study dynamics of cascades in Section~\ref{sec:analysis}.
We illustrate the framework on simple cascades often observed in online social networks. In Section~\ref{sec:digg} we also apply it to study large information cascades occurring on a real-world social network of Digg (http://digg.com). This site allows people to submit and vote for news stories, and also to create links to other people in order to see what new stories they have recently voted for. Stories propagate on Digg's social network through a series of cascades as users influence their fans to vote for the story~\cite{Lerman10icwsm, Ghosh10}. We study the distribution of several macroscopic properties of these cascades.
In addition, we study the microscopic dynamics of their temporal evolution. Time plots of the cascade generating function show several characteristic signatures of cascade growth, such as star-like, chain-like and community-like growth.

\section{A Framework for Analyzing Cascades}
\label{sec:framework}

\label{sec:definitions}
\begin{figure}[tbh]
\scalebox{0.7}
{
\begin{tabular}{@{}|c@{}|@{}c|@{}}
\hline
 \includegraphics[width=0.65\columnwidth]{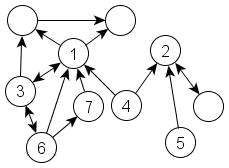} &
 \includegraphics[width=0.65\columnwidth]{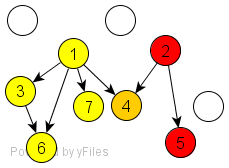}\\
 (a) & (b) \\  \hline
\end{tabular}
}
  \caption{An toy example of an information cascade on a network. Nodes are labeled in the temporal order in which they are activated by the cascade.  The nodes that are never activated are blank. (a)  The edges show the underlying friendship network. Edge direction shows the semantics of the connection, i.e., nodes are watching others to which they are pointing. (b) Two cascades on the network (shown in yellow and red).   Node 1 is the seed of the first (yellow) cascade and node 2 is the seed of the second (red) cascade. Node 4 belongs to both cascades and is shown in orange. }
  \label{fig:network}
\end{figure}
Consider a social network, represented by a graph $G(V,E)$ with $V (|V|=N)$ nodes and $E(|E|=M)$ directed edges.
If node $a$ wants to watch activities of node $b$, she must create an edge to $b$ by designating $b$ as a $friend$. We call $a$ a \emph{fan} (or  {follower}) of $b$. Figure~\ref{fig:network}(a) shows a directed network in which node $4$ is a fan of $1$ and $2$. We call an edge $e_{ij}$  \emph{active}, if node $j$ is a fan of node $i$ and node $i$ is {activated} before node $j$. Information or influence flows \emph{from} activated nodes \emph{to} their fans. In the figure above, information flows from nodes $1$ and $2$ {to} $4$.

A \emph{cascade} is a sequence of activations generated by a \emph{contagion process}, in which nodes cause connected nodes to be activated with some probability. In analogy with the spread of an infectious disease on a network, an \emph{infected} (activated) node \emph{exposes} his fans to the infection. Disease cascades through the network as exposed fans become infected, thereby exposing their own fans to the disease, and so on. The \emph{seed} of a cascade is the node that initiates the cascade.  In information cascades the seed is an independent originator of information, who then influences others to adopt, endorse, or transmit that information. We call a node that participates in a cascade a \emph{member} of the cascade. A contagion process can generate multiple cascades, and a node can participate in more than one cascade, resulting in a commonly observed ``collision of cascades''~\cite{Leskovec07} phenomenon. Figure~\ref{fig:network}(b) shows cascades on the network shown in Fig.~\ref{fig:network}(a), in which nodes are labeled in the order they are activated, with links showing the direction of influence. As shown, the contagion process generates two cascades whose seeds are nodes $1$ and $2$, respectively. Node $4$ participates in both cascades.

A \emph{cascade chain} is a sequence of connected nodes participating in a cascade. Each node in the {cascade chain} is influenced by all the nodes in the chain activated before it and influences  all the successive nodes in the chain. The length of the longest  chain is the \emph{diameter} of the cascade~\cite{Leskovec07}.  The \emph{spread} of the cascade is the maximal branching number of its participants, i.e., the maximum number of nodes a single member infects. The diameter of the contagion process in Fig. \ref{fig:network}(b) is two (longest chain is $1 \rightarrow 3 \rightarrow 6$, the spread of cascade 1 (yellow) is 4 and of cascade 2 (in red) is 2.

\subsection{Characterizing Cascades}
\label{sec:characterization}
We characterize a {cascade} mathematically by the \emph{cascade generating function},  $\phi(j,\alpha_{j,i})$, which describes how activation spreads through the network.  Contagion process is parameterized by the transmission rates $\alpha_{j,i}$  $ \forall j, i  \in [1,N]$, which  give the probability that a node $i$ activated at time ${t_i}$ will activate a connected node  $j$ at a later time $t_j$. Though, in principle, $\alpha_{j,i}$ could be different for different values of $i$ and $j$, for simplicity, we assume that they are all the same, i.e., $\alpha_{ji}=\alpha$. Note, that since the nodes are labeled in the temporal order of their activation, $\phi(j,\alpha_{j,i})$ characterizes the cascade at time $t_j$.


We use the contagion process shown in Fig.~\ref{fig:network}(b) to illustrate how the cascade generating function is calculated.
The initial value of the cascade function is some constant. In the example, nodes $1$ and $2$ are seeds; therefore, the values of the cascade function at the times they are activated are constant. While these values may be different, for convenience we set them both to one: $\phi(1,\alpha)=\phi(2,\alpha)=1$. The value of $\phi$ captures the cumulative effect on node $j$ of activated nodes that are connected to $j$. Node $3$ is connected to $1$ and activated by it with probability $\alpha$; therefore, $\phi(3, \alpha)=\alpha \phi(1, \alpha)$. At the time node $4$ is activated, cascade function is $\phi(4, \alpha)=\alpha \phi(1,\alpha) + \alpha \phi(2,\alpha)$. Nodes continue to activate others in this fashion. At time $t_6$, the cascade function is $\phi(6,\alpha)=\alpha \phi(1,\alpha)+\alpha \phi(3,\alpha)$. Since $\phi(3,\alpha)$ only depends on $\phi(1,\alpha)$, $\phi(6, \alpha)$ can be rewritten as
$ \phi(6,\alpha)=\alpha \phi(1, \alpha)+\alpha^2 \phi(1,\alpha)$.

In general terms, if  node $i$ is a node activated at time $t_i$, the value of the cascade generating function at later time $t_j$ when node $j$ is activated is:
 \begin{equation}
\phi(j, \alpha )=\sum_{i \in friend(j)}  \alpha \phi(i, \alpha)
\label{eq:phi}
\end{equation}
\noindent
where $friend(j)$ is a set of nodes connected to node $j$ that are activated before it. Since links are directed, without loss of generality, we can assume that there are $K$ cascades in a contagion process. Let $\phi(i_1,\alpha)$, $\phi(i_2,\alpha)$, $\cdots$, $\phi(i_K,\alpha)$ be the weights of their seeds.
 Then, Eq.~\ref{eq:phi}  reduces to
 \begin{equation}
\phi(j, \alpha )=\sum_{i \in friend(j)}\alpha \phi(i, \alpha)=  \sum_{p=1}^{K}  f(j, i_p, \alpha) \phi(i_p, \alpha)
\label{eq:phi1}
\end{equation}
\noindent The value of $\phi(j, \alpha) $ is proportional to the cumulative effect or influence of all cascades on node $j$ activated at time $t_j$ and can be described using the vector $(f(j, i_1, \alpha),\cdots,f(j, i_K, \alpha) )$.
$f(j, i_p, \alpha),$ captures the cumulative effect  of the cascade generated at seed node $i_p$ on the node $j$ where $t_j> t_{i_p} $. In Fig.~\ref{fig:network}(b), at time $t=4$, $\phi(4,\alpha)= f(4,1, \alpha)\phi(1,\alpha)+f(4,2, \alpha)$
 $\phi(2,\alpha)$ where  $f(4,1, \alpha) =\alpha$ and $f(4,2,\alpha)=\alpha$. At time  $t=6$, $\phi(6,\alpha)= f(6,1, \alpha)\phi(1,\alpha)+ f(6,2, \alpha)\phi(2,\alpha)$. Here $f(6,1, \alpha)=\alpha+\alpha^2$ and $f(6,2, \alpha)=0$.

If the values of the cascade generating function for nodes $i$ and $j$ are the same, $\phi(i,\alpha)=\phi(j,\alpha)$,
the nodes $i$ and $j$ are \emph{ isomorphic} with respect to the contagion process. Such nodes are \emph{structurally similar} with respect to the cascade; therefore, the value of the cascade function is independent of the order in which they are activated. By structural similarity, we mean that in a network comprising of only the activated nodes and active edges between them,  the topological distance of two isomorphic  from all the seeds is the same. Here, the topological distance of a node from the seed is measured in terms of the total number of attenuated paths over active edges.
Isomorphic nodes can be grouped together in a \emph{tier} with its own {characteristic} $\phi(\alpha)$. In the contagion process in Fig.~\ref{fig:network}(b), nodes $3$ and $7$ are isomorphic and form a tier with value $\phi(\alpha)=\alpha \phi(1,\alpha)$.

\textbf{Cascade properties.}
We can use the cascade generating function to compute the macroscopic properties of cascades, such as their size, diameter, number of paths, and their average length.

If we take $\phi(i_p,\alpha)=1$,  where $i_p$ is the seed of $p^{th}$  cascade activated at time $t_{i_p}$, then the {total number} of paths  from $i_p$ to node $j$ is equal to  $f(j, i_p, 1)$  in Eq.~\ref{eq:phi1}.
The {total length} of paths  from the seed ${i_p}$ to $j$, $ l(j, i_p)$, can be obtained by  differentiating $\phi$ with respect to $\alpha$ and evaluating the derivative at $\alpha=1$ i.e $ l(j, i_p) =\frac{df(j, i_p, \alpha)}{d \alpha} |_{\alpha=1}$
\begin{equation}
\frac{d \phi(j,\alpha)}{d\alpha}|_{\alpha=1} = \sum_{p=1}^{K}  \frac{df(j, i_p, \alpha)}{d \alpha} \phi(i_p,\alpha) 
 =  \sum_{p=1}^{K}  l(j, i_p) \phi(i_p, \alpha)
\label{eq:len}
\end{equation}
\noindent To illustrate this, consider again the contagion process shown in Fig.~\ref{fig:network}(b). For example, if we pick node $6$, there are two paths from the seed (node $1$)  to node $6$: $1 \rightarrow 3 \rightarrow 6$ and $1 \rightarrow 6$.  The total length of these paths is three. There are no paths from the second seed (node 2 ) to node $6$.  We can also get this answer from
$$\frac{d\phi(6,\alpha)}{d\alpha}|_{\alpha=1}=\frac{d(\alpha +\alpha^2)\phi(1,\alpha) }{d \alpha}=3.$$


We can use similar reasoning to compute other cascade properties. The  {average path length}, $l_{av}$   is given by:
\begin{equation}
l_{av}= \frac{\sum_{j} \sum_{p=1}^{K}  l(j,i_p)}{ \sum_{j}\sum_{p=1}^{K}  f(j,i_p,1)}= \frac{d \sum_j  \phi(j,\alpha)}{d\alpha}\frac{\alpha}{\sum_j \phi(j,\alpha)}|_{\alpha=1}
\label{eq:av}
\end{equation}
\noindent
The diameter of the contagion process is the length of the longest path of any cascade generated by this process. It is given by  modifying Eq.~\ref{eq:av}:
 \begin{equation}
l_{max}= \max_{j \in [2,N]}{\frac{d \phi_{min}(j,\alpha)}{d\alpha}\frac{ \alpha}{\phi_{min}(j,\alpha)}},
\label{eq:diam}
\end{equation}
\noindent where
\begin{equation*}
\phi_{min}(j, \alpha )=\min_{i \in friend(j)} \alpha \phi_{min}(i, \alpha)
\label{eq:phi2}
\end{equation*}

\begin{figure*} [!htbp]
\begin{center}
  \begin{tabular}{@{}c|@{}c@{}|@{}c@{}|@{}c@{}|@{}c|@{}c@{}|@{}c@{}|@{}c@{}| }
\multicolumn{1}{c}{ }
& \multicolumn{1}{c}{(1)}
 &  \multicolumn{1}{c}{(2)}
 & \multicolumn{1}{c}{(3)}
  &  \multicolumn{1}{c}{(4)}
 & \multicolumn{1}{c}{(5)}
  &  \multicolumn{1}{c}{(6)}
    &  \multicolumn{1}{c}{(7)}\\
\cline{2-8}{(a)} &
\includegraphics[width=0.9in]{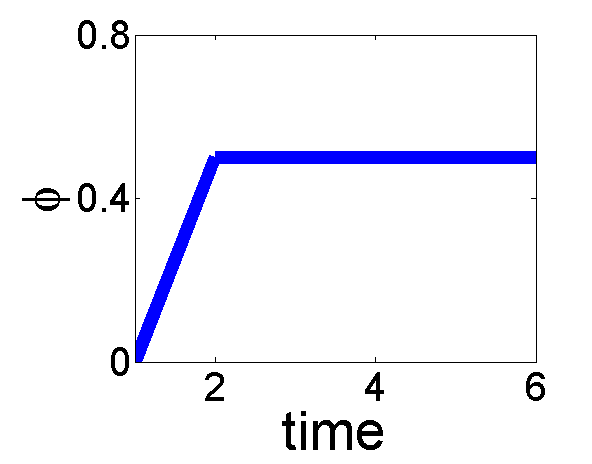} &
\includegraphics[width=0.9in]{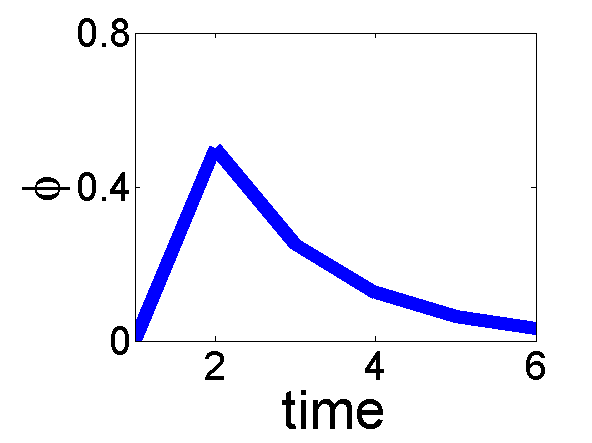}  &
\includegraphics[width=0.9in]{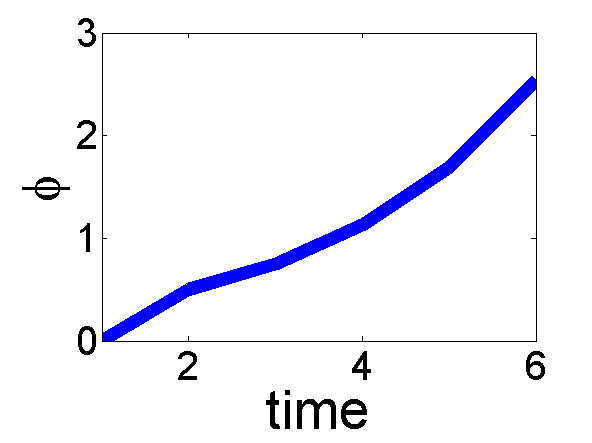} &
\includegraphics[width=0.9in]{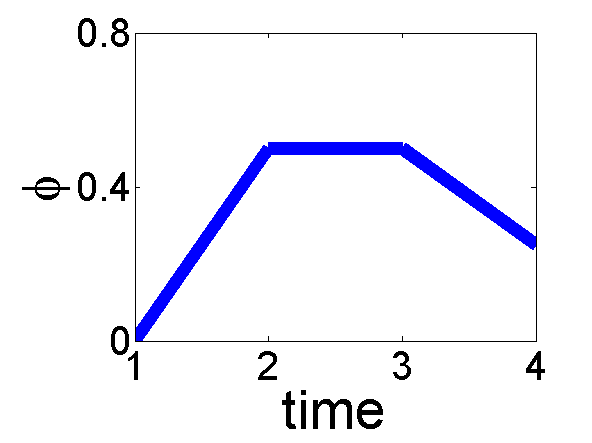} &
\includegraphics  [width=0.9in]{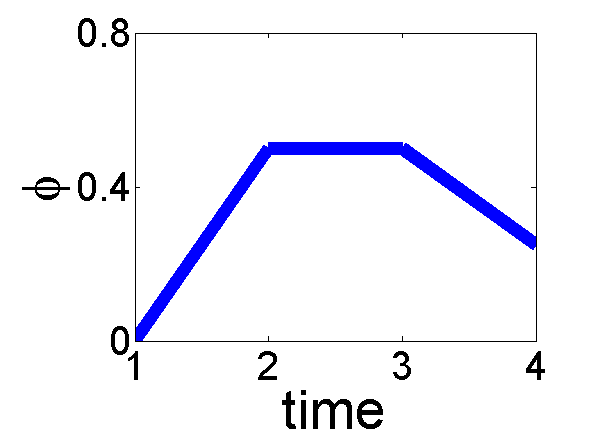} &
\includegraphics  [width=0.9in]{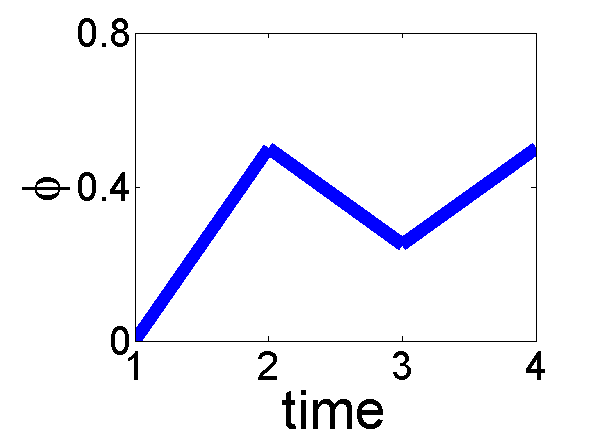} &
 \includegraphics  [width=0.9in]{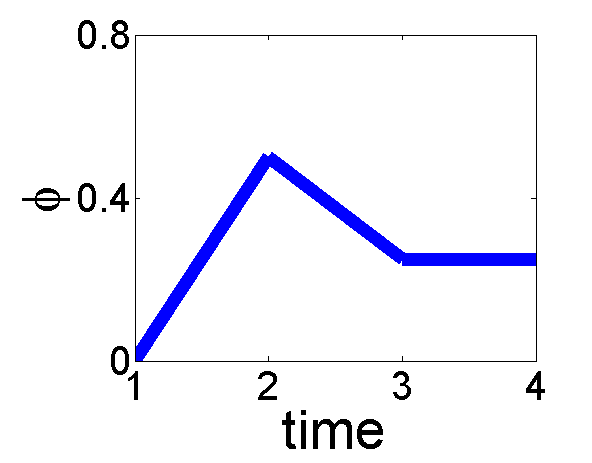} \\
\cline{2-8}
(b)& \includegraphics [width=0.9in,height=0.9in]{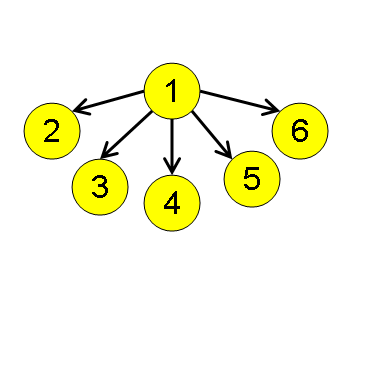}&
 \includegraphics [width=0.9in,height=0.9in]{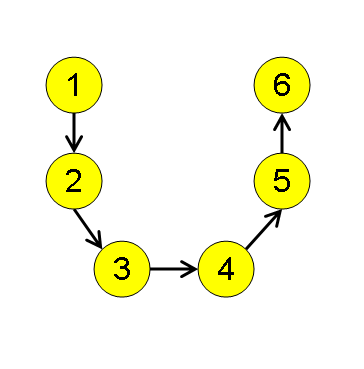}&
 \includegraphics [width=0.9in,height=0.9in]{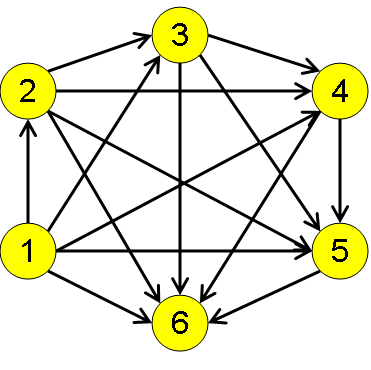}&
 \includegraphics [width=0.9in,height=0.9in]{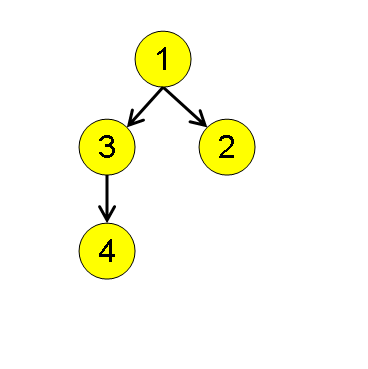}&
 \includegraphics [width=0.9in,height=0.9in]{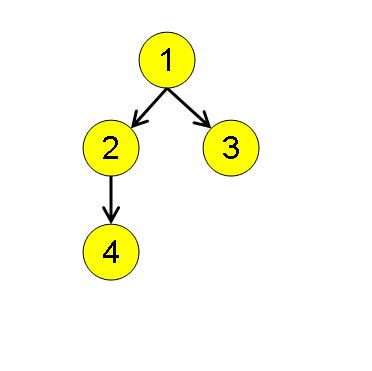}  &
 \includegraphics [width=0.9in,height=0.9in]{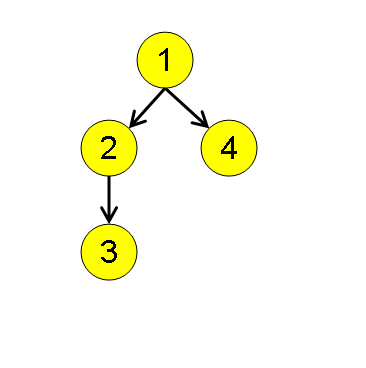} &
 \includegraphics [width=0.9in,height=0.9in]{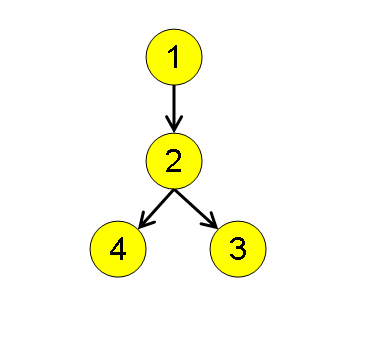}\\
\cline{2-8}
&{tot.paths=5}&{tot.paths=5}&{tot.paths=31}&{tot.paths=3}&{tot.paths=3}&{tot.paths=3}&{tot.paths=3} \\
&{tot.len=5}&{tot.len=15}&{ tot.len=80 }&{tot.len=4}&{tot.len=4}&{ tot.len=4}&{ tot.len=5} \\
&{av.len=1}&{av.len=3}&{ av.len=2.58}&{av.len=1.33}&{av.len=1.33}&{av.len=1.33}&{ av.len=1.67} \\
{(c)}&{diam.=2 }&{diam.=5}&{diam.=5}&{diam.=2 }&{ diam.=2}&{ diam.=2 }&{diam.=2} \\
\cline{2-8}
&{\{1\},\{2,3,4,5,6\}}&{\{1\},\{2\},\{3\},}&\large{\{1\},\{2\},\{3\}, }&{ \{1\},\{2,3\},\{4\} }&{ \{1\},\{2,3\},\{4\} }&{ \{1\},\{2,4\},\{3\}}&{ \{1\},\{2\},\{3,4\}} \\
{(d)  }&                   &{\{4\},\{5\},\{6\} }&{ \{4\},\{5\},\{6\} } & &&&\\
\cline{2-8}
\\
\cline{2-8}
 (a) &
\includegraphics  [width=0.9in]{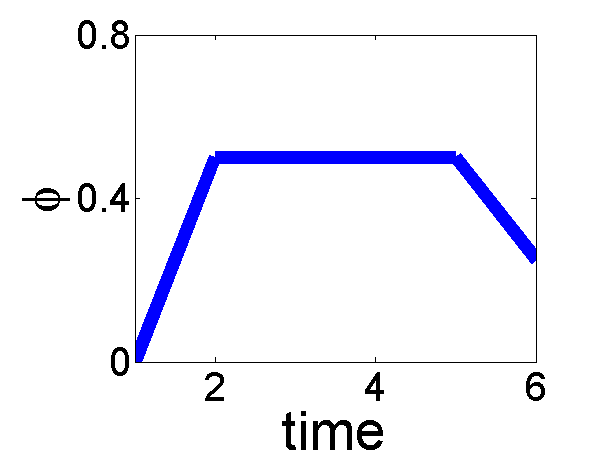}  &
\includegraphics  [width=0.9in]{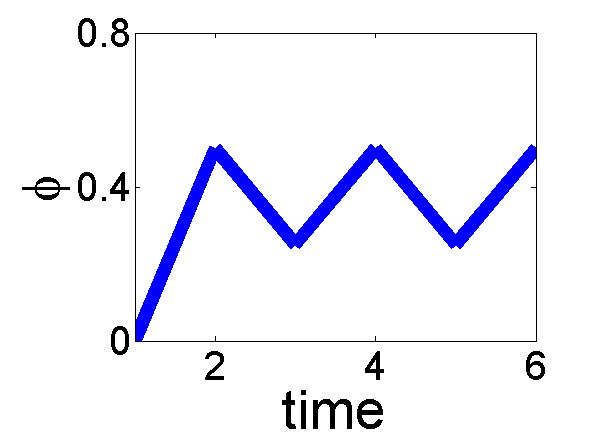} &
\includegraphics  [width=0.9in]{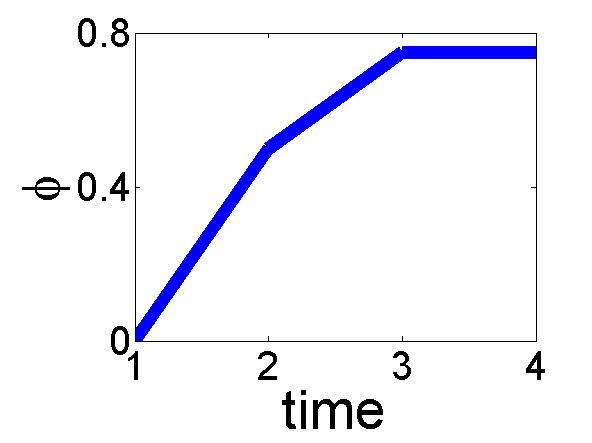} &
\includegraphics  [width=0.9in]{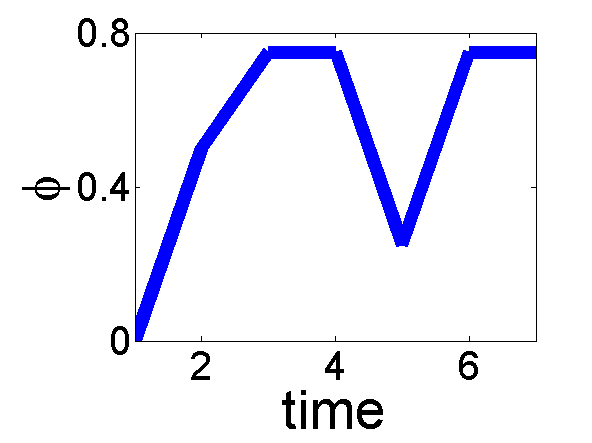}  &
\includegraphics  [width=0.9in]{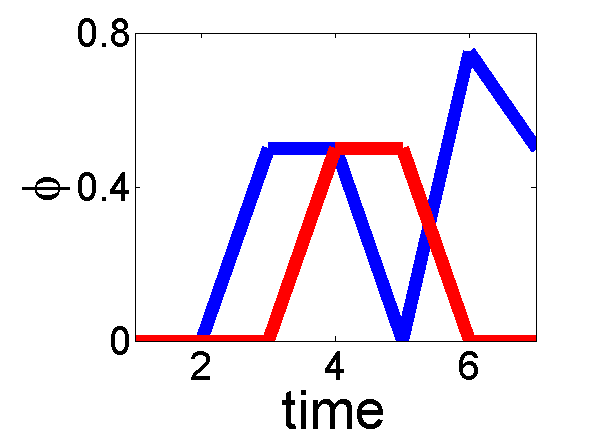} &
\includegraphics  [width=0.9in]{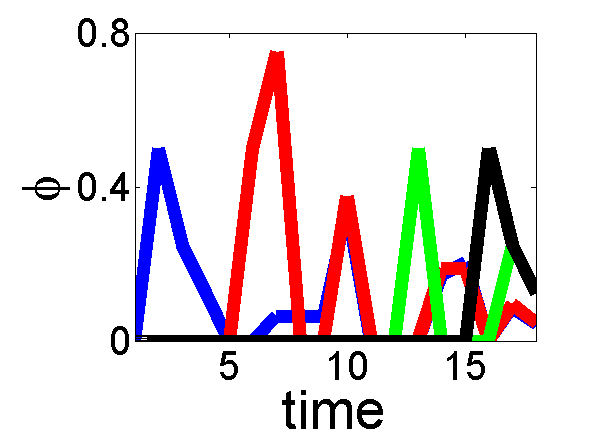}  &
\includegraphics  [width=0.9in]{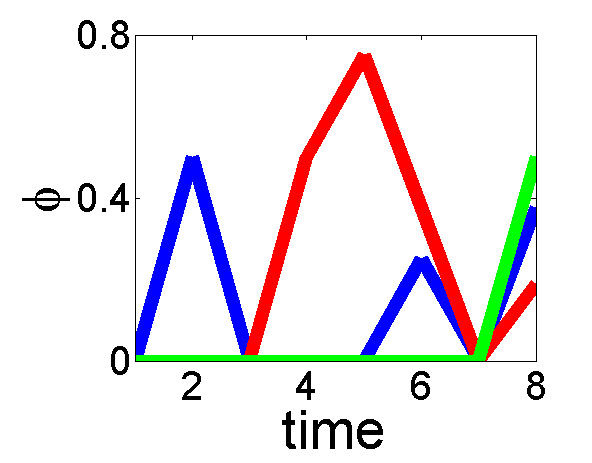}  \\
\cline{2-8}
(b) &
 \includegraphics [width=0.9in,height=0.9in]{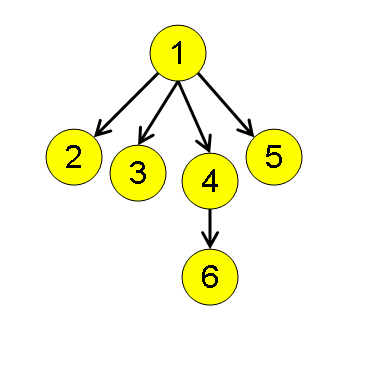}&
 \includegraphics [width=0.9in,height=0.9in]{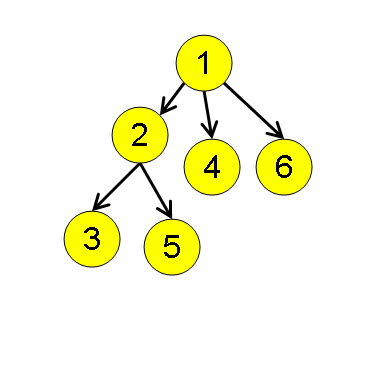}&
 \includegraphics [width=0.9in,height=0.9in]{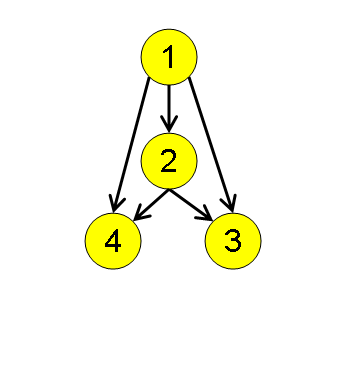}&
 \includegraphics [width=0.9in,height=0.9in]{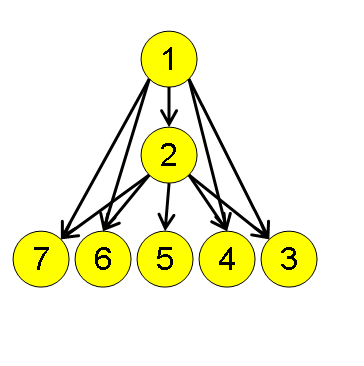}&
 \includegraphics [width=0.9in,height=0.9in]{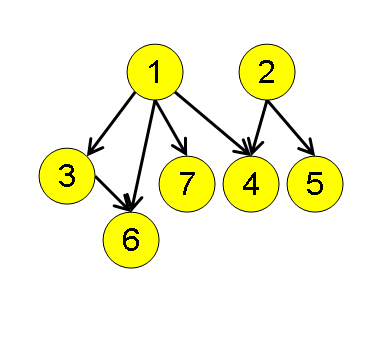} &
 \includegraphics [width=0.9in,height=0.9in]{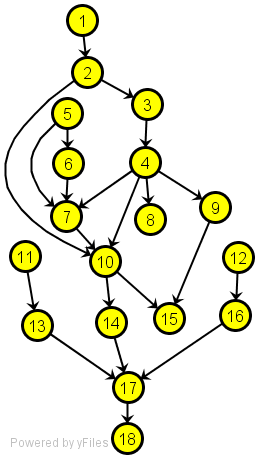} &
 \includegraphics [width=0.9in,height=0.9in]{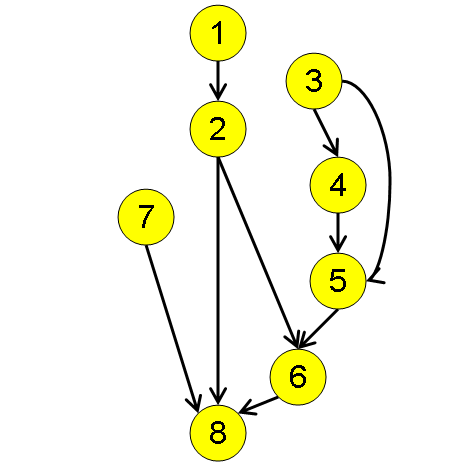}  \\
\cline{2-8}
 & \large{tot.paths=5} & \large{tot.paths=5}& \large{tot.paths=5}& tot.paths=10& tot.paths=7& tot.paths=41& tot.paths=12\\
&tot.len=6 &tot.len=7&tot.len=7&tot.len=15&tot.len=8 &tot.len=154&tot.len=25\\
& av.len=1.2&av.len=1.4&av.len=1.4&av.len=1.50&av.len= 1.14&av.len=3.76&av.len=2.08\\
(c)& diam.=2    &diam.=2&diam.=2&diam.=2&diam.=2 &diam.=8      &diam.=4\\
\cline{2-8}
   & \{1\},\{2,3,4,5\}&\{1\},\{2,4,6\}&\{1\},\{2\}&\{1\},\{2\},&\{1,2\},\{3,7\},&\{1,5,11,12\}, $\ldots$ & \{1,3,7\},\{2\},\{4\}, \\
(d)& \{6\}           &\{3,5\}  &  \{3,4\}         & \{3,4,6,7\},\{5\} &\{5\},\{4\},\{6\}                &\{8,9\},$\ldots$,\{18\} & \{5\},\{6\},\{8\} \\
 \cline{2-8}
\multicolumn{1}{c}{ }
& \multicolumn{1}{c}{(8)}
 & \multicolumn{1}{c}{(9)}
  &  \multicolumn{1}{c}{(10)}
 & \multicolumn{1}{c}{(11)}
  &  \multicolumn{1}{c}{(12)}
  &  \multicolumn{1}{c}{(13)}
&  \multicolumn{1}{c}{(14)} \\
  \end{tabular}
       \end{center}
  \caption {Analysis of various cascades. Nodes are labeled by the order in which they are activated by the contagion process. Row (a) shows cascade plots obtained by computing the cascade generating function, $\phi$ at different times. Row (b) shows the corresponding contagion process. Different cascades within the same contagion process are shown in different colors. Row (c) shows some of the numeric properties of the cascades, and row (d) shows sets of isomorphic nodes.}
 \label{cascade_fig}
  \end{figure*}

\subsection{Computational Framework for $\phi$}
\label{sec:comp}

\paragraph{Cascade Graph}
For the analysis of the contagion process, we create a \emph{cascade graph} $G_c(V_c,E_c)$  from the original network $G(V,E)$ as follows. Let $V_c$ be the number of nodes participating in all cascades. Let a cascade begin at time $t_1$ and end at $t_N$. We arrange  and label the nodes in the temporal order in which they are activated, e.g., transmit information: $1,2,\cdots, N$, where node $k$ activated at time $t_{k}$ and  $t_1\le \cdots  \le t_k \le \cdots \le t_N$.
An edge exists from $j$ to $i$ in $G_c$ (i.e. $i$ is activated by $j$) if an edge exists from $i$ to $j$ in $G$ ($i$ is a fan of $j$) and $t_i > t_j$. The adjacency matrix of $A_c$ of  the cascade graph $G_c(V_c,E_c)$, the \emph{cascade matrix},  is:
\begin{eqnarray*}
A_c(i,j) &= &1 \mbox{  if $\exists$ an edge from $j$ to $i$ in $G_c(V_c,E_c)$ and $j<i$} \\
&=& 0 \mbox{ otherwise}
\end{eqnarray*}
\noindent We break ties randomly. If nodes $a$ and $b$ receive information at the same time $t_k$, without loss of generality, we assume $a=k$ and $b=k+1$. Also, we modify the adjacency matrix $A_c$, making $A_c(k+1,k)=0$ and $A_c(k,k+1)=0$, irrespective of  whether or not an edge exists between $k$ and $k+1$. This means that neither node can activate the  other,  since they are activated at the same time.
We note that $1$ is always the {seed} of a cascade. The cascade matrix can encode a contagion process that generates multiple cascades.

\paragraph{Contagion and Length Matrix}
In addition to the cascade matrix, we introduce the \emph{dynamic adjacency matrix} of the cascade graph, $A(t)$. This is a time-dependent matrix, whose non-zero elements include all nodes that have been activated up to time $t$:
\begin{eqnarray}
A_{ij}(t_k) & = & 1 \mbox{ \emph{ if  $A_c(i,j)=1$ and $t_i\le t_k$ }} \nonumber \\
&=& 0 \mbox{ \emph{otherwise}} \nonumber
\label{eq_1}
\end{eqnarray}

The dynamic adjacency matrix allows us to compute connectivity between nodes in a cascade, as measured by the number of paths that exist between them. Following \cite{Lerman10mlg}, let the attenuation parameter $\alpha$ be the probability of transmitting a message or influence along any edge from node $i$ at time $t_i$ to node $j$ at time $t_j$.
The \emph{contagion matrix}  over the time period $[t_1, t_N]$:
\begin{eqnarray}
C({\alpha}) &=& \alpha^{N-1}A({t_N})\cdots  A({t_3})A({t_2}) +\cdots\nonumber \\
    &+&\alpha^{2} A({t_N})A({t_{N-1}})+\alpha A({t_N})+I
 \label{eq:contagion_mat}
\end{eqnarray}
\noindent The term $C_{ij}({\alpha})$ gives the number of attenuated paths from node $j$ to $i$ in $G_c(V_c,E_c)$ and $I$ is the identity matrix.


The total length of paths from one node to another can be modeled using a formalism similar to contagion matrix.
We  define the \emph{length matrix} as:
\begin{eqnarray}
L(\alpha) &=& (N-1)\alpha^{N-1}A({t_N})\cdots A({t_3})A({t_2}) \\ \nonumber
    &+&\cdots+2\alpha^{2} A({t_N})A({t_{N-1}})+\alpha A({t_N})+I
 \label{eq:path_mat}
\end{eqnarray}
\noindent
Here $L_{ij}(1)$  gives the total length of paths from node $j$ to node $i$ in $G_c(V_c,E_c)$.
${L_{ij}(1)}/{C_{ij}(1)}$ then gives the \emph{average length} of paths from node $j$ to node $i$.

The first step towards quantifying cascades is seed identification. The  can be achieved by collecting all the maximal elements of $G_c$, seen as partially ordered set. Equivalently, if  all the elements of the $i^{th}$ row of $A_c$ are zero, then node $i$ is a {seed} of the cascade.  Finding all the seeds gives the total number of cascades, $K$, in the contagion process. Let $\phi(i_1,\alpha),\cdots,\phi(i_K,\alpha)$ be the cascade function value of each seed. The value of the cascade function of node $j$, which is activated after $m$ (but before $k-m$) cascade seeds, is $\phi(j, \alpha)=\sum_{p=1} ^{m} C_{j,i_p}(\alpha)\phi(i_p,\alpha)$. A non-zero value of $C_{j,i_p}(\alpha)$ indicates that $j$ is a member of the cascade initiated by ${i_p}$. Hence, $C_{j,i_p}(\alpha)=f(j, i_p, \alpha)$  in Eq. \ref{eq:phi1}. The efficient design of the cascade generating function is such, that knowing only the columns corresponding to the seeds in the contagion matrix, would help us to characterize the entire matrix.  This along with the triangular nature of the cascade matrix enables us to calculate the contagion and length matrices, and hence the corresponding cascade generating function efficiently using dynamic programming (Algorithm \ref{alg:2} ). This algorithm has  $O(kN)$ space and $O(dkN)$ runtime complexity even in its naive implementation, where, $d$ is the maximum degree of any node and $N$ is the number of nodes in the process.
{\small
{
\begin{algorithm}
\caption{Efficient algorithm for computing the Contagion and the Length Matrix}
\label{alg:2}
\begin{algorithmic}
\small
{
\STATE{\bf{Input}}\\
$A_c$: Adjacency Matrix of the Cascade Graph  \\
$\alpha$: transmission probability \\
\STATE{\bf{Output}}\\
$C(\alpha)$: Contagion Matrix ($N \times k$), $L(\alpha)$: Length Matrix ($N \times k$) \\
$p^{th}$ column in C and L, corresponds features of the cascade generated by the $p^{th}$ seed.\\
$\forall p \in [1,k]$,  $j$ is the label of  the $p^{th}$ seed  activated at time $t_j$.\\
$C_{i,p}(\alpha)$ is the cascade generating value for node $i$ with respect to the $p^{th}$ seed. $L_{i,p}(\alpha)$ at $\alpha=1$ gives the total length of paths from the $p^{th}$ seed to the node $i$.
$\forall p C_{j,p}(\alpha)=1, L_{j,p}(\alpha)=1 $\\
\IF {$i < j$}
        \STATE $C_{i,p}(\alpha)=0, L_{i,p}(\alpha)=0$
\ELSE
        \IF {$i==j+1$}
                \STATE $C_{i,p}(\alpha)=\alpha A_c(i,j), L_{i,p}(\alpha)=\alpha A_c(i,j)$
         \ELSE
                \STATE $C_{i,p}(\alpha)=\alpha A_c(i,j)+ \displaystyle \sum_ {\forall edges\ e(i-k,i)| k=1}^{i-j-1 }\alpha A_c(i,i-k)C_{i-k,p}(\alpha)  $
                 \STATE$L_{i,p}(\alpha)=\alpha A_c(i,j)+ \displaystyle \sum_ {\forall edges\ e(i-k,i)| k=1}^{i-j-1 } \alpha A_c(i,i-k) ( C_{i-k,p}(\alpha)+ L_{i-k,p}(\alpha))  $

        \ENDIF
\ENDIF
}
\end{algorithmic}
\end{algorithm}
}
}

The \emph{contagion} and \emph{length} matrices together fully determine $\phi$ and  $\frac{d\phi}{d\alpha}$, and therefore, capture the microscopic details of the contagion process.
If vector $\mathbf{c}=( C_{j_1,i_1}(\alpha)$,  $C_{j_1,i_2}(\alpha)$, $\cdots,  C_{j_1,i_K}(\alpha))$ $=( C_{j_2,i_1}(\alpha)$, $C_{j_2,i_2}(\alpha)$, $\cdots$, $C_{j_2,i_K}(\alpha))$, then $j_1$ and $j_2$ are \emph{isomorphic} with respect to the contagion process.

The  {total number} and {total length} of paths in the cascade from seed ${i_p}$ to node $j$  is given by $C_{j,i_p}(1) =f(j, i_p, 1)$  and  $  L_{j,i_p}(1)= l(j, i_p)$  in Eq.~\ref{eq:phi1} and Eq.~\ref{eq:len} . Hence, the {total number of paths}, {total length}, and {average path length}  for the entire contagion process is given by  $\sum_{i_p}\sum_{j\neq i_p \forall p} C_{j,i_p}(1)$,  $\sum_{i_p}\sum_{j\neq i_p \forall p} L_{j,i_p}(1)$ and $\frac{\sum_{i_p}\sum_{j\neq i_p \forall p} L_{j,i_p}(1)}{\sum_{i_p}\sum_{j\neq i_p \forall p} C_{j,i_p}(1)}$.

As can be seen in Eq. \ref{eq:diam}, analogous to the the length matrix, we have devised an efficient algorithm to calculate the diameter. Due to lack of space, we do not provide the algorithm here. Since its formalism is very similar to that of the  length matrix, computation has comparable runtime  and space complexity.

\subsection{Analyzing Cascades}
\label{sec:analysis}
Plotting the cascade generating function  $\phi(j,\alpha)$ \emph{vs} time ($j$) shows how the structure of the cascade evolves over time. Fig.~\ref{cascade_fig} illustrates the cascade plots computed for a variety of contagion processes, which include several prototypes of cascades frequently observed in recommendation and blog networks~\cite{Leskovec07, Leskovec05}. We label nodes in the order in which they are activated and take $\phi(i,\alpha)=1$ when $i$ is the seed of the cascade, thus giving equal  weights to all cascades in the contagion process.
 Without loss of generality, in this study, we set the value of $\alpha$ to 0.5. Future work includes estimation of the transmission rate empirically from the network. We show that cascade plots contain as much information as cascade graphs, but can be used to analyze the structure and evolution of even large cascades, for which visualization is not feasible. In addition to showing the cascade plot for each cascade (row (a)), Fig.~\ref{cascade_fig} also reports some of the macroscopic properties of the cascade (row (c)), such as total number of paths and their length, average path length, and diameter. Note that this is not the exhaustive list of properties that can be calculated using the cascade characterization function. Row (d) lists groups of isomorphic nodes in each cascade.

Cascades (1)--(3) in Fig.~\ref{cascade_fig} are three of the commonly observed patterns, such as a star (Fig.~\ref{cascade_fig}(1)), a chain (Fig.~\ref{cascade_fig}(2)), and a community (clique) (Fig.~\ref{cascade_fig}(3)).  In the star-like contagion process, Fig.~\ref{cascade_fig}(1), nodes activated by $n_1$ have the same value of $\phi$, and form an isomorphic group. Interchanging the order of their activation does not affect the value of $\phi$ or the cascade plot. In the chain-like contagion process, cascade function decreases as the chain becomes longer. There are no isomorphic nodes. In the clique-like contagion process, the value of the cascade function grows in time as more paths are created in the cascade. There are also no isomorphic nodes in this cascade.

In the contagion process in Fig.~\ref{cascade_fig}(4), nodes  activated at $t=2$ and $t=3$ are  { isomorphic}, therefore, the evolution of this cascade is indistinguishable from the cascade shown in Fig.~\ref{cascade_fig}(5). However, if the shape of the cascade is the same, but nodes are activated in different order, Fig.~\ref{cascade_fig}(6), the cascade plot and its structure are different. This is because in the contagion processes (4) and (5), cascade widens first (it is star-like), before lengthening, while in the contagion process (6), cascade lengthens first (it is chain-like), before widening. Similarly, the cascade (7) first deepens, then widens, opposite of cascade (8), while cascade (9) alternates between deepening and widening. In none of these cascades (except (3)) are there multiple paths to a node. Once this happens, as in cascades (10) and (11), the value of the cascade function increases.

We can also disentangle multiple cascades co-occurring in a contagion process. Contagion processes (12)--(14) contain multiple cascades, whose cascade functions are shown in different color. Note that in the contagion process (12), node $4$ is isomorphic to $3$ and $7$ with respect to the cascade initiated by $1$, and it is isomorphic to $5$ with respect to the cascade initiated by $2$.

\subsection{Reconstructing Cascades}
\label{sec:reconstructing}
Given the {contagion matrix}, it is possible to reconstruct the contagion process with a high level of accuracy. The  {cascade generating function} $\phi$, compresses information, and has a space complexity of $O(KN)$ where $K$ is the number of seeds in the contagion process. How well does the compressed representation $\phi$ capture the contagion process?

Using $\phi$ with $0 < \alpha <1$, a tier-level reconstruction of the contagion process is possible. This {reconstruction} does not remove degeneracy of {isomorphic} nodes.   Temporal ordering of the nodes, help us to {fine-tune} the {tier-level reconstruction}. Additional information, such us the number of nodes $m$ and their $indegree$ and $outdegree$ can help us further improve the approximation. In all the examples shown in Fig.~\ref{cascade_fig},   using just $\phi$, we are  able to  obtain the \emph{exact  tier-level reconstruction} in all cases. To illustrate, consider Fig.~\ref{fig:network}, taking $\phi(1)=\phi(2)=1$,
we get $\phi(3,\alpha)=\phi(7, \alpha)=(\alpha,0)$, where $\alpha$ is the value of the cascade function for the cascade initiated by seed $1$, and $0$ is the value of the cascade function for the cascade initiated by second seed, node $2$. Likewise, $\phi(4,\alpha)=(\alpha,\alpha)$, $\phi(5,\alpha)=(0,\alpha)$ and $\phi(6,\alpha)=(\alpha+\alpha^2,0)$. Hence we can reconstruct that nodes 1 and 2 are independent seeds, 3 and 7 are connected to only node 1 and 5 is connected to only node 2. Node 4 is connected to both 1 and 2. Node 6 is connected to  1 and to that \emph{tier} of nodes containing node 3 and 7. However due to the temporal arrangement of the nodes, we know that 6 is activated before 7, hence it is necessarily connected to node 3. Thus we are able to obtain the \emph{exact} reconstruction of the cascade.

In Fig.~\ref{cascade_fig},  using just $\phi$, we are able to obtain the \emph{exact  tier-level reconstruction} for cases 1, 4, 5, 7, 8, 9,10, 11 and 13. In most  of these cases we are also able to disambiguate between isomorphic nodes in the same tier. For cases 2, 3, 6,  12, and 14, we are also able to obtain \emph{exact node-level reconstruction} of the cascade graph.

\paragraph{Space and time complexity}
Clearly, as demonstrated by the discussion above, knowing the values of $\phi$ at different times allows us to deduce the dynamics of a cascade, and reconstruct its structure (up to the degeneracy that exists for isomorphic nodes). Storing the shape of the cascade has $O(N^2)$ space complexity. However, as demonstrated above, the cascade generating function can reconstruct this shape with high degree of accuracy. Having a pseudo-linear space complexity $O(KN)$, it provides an efficient  compression of this information.
Besides, this model is general, because the same model can be used to investigate cascades in information flow, epidemics, computer viruses, and so on.
This method is fast having $O(dKN)$ runtime complexity even in its naive implementation where $d$ is the maximum degree of any node. Moreover, the cascade generating function of a node activated at time $t$, depends only on the cascade value of his friends activated before him.  Hence $\phi$ can be calculated real-time and is appropriate even for applications which require streaming, online or near real-time analysis of cascades.

\section{Digg Case Study}
\label{sec:digg}
We use the framework described above to study information spread on the social news aggregator Digg which allows users to \emph{post} and \emph{vote for} news stories. Digg users can also create  social networks by adding others as {friends}.
Digg highlights the stories a user's friends posted or voted for by marking them with a green ribbon and also displaying them on the \emph{Friends Interface}, a special page for watching friends' activity.
 A fan may then see the story, and if she decides to vote for it, the story then becomes visible to her own fans, who may in turn vote for it, etc. By voting for a story, a user may influence her fans to also vote it~\cite{Ghosh10}. The spread of a story through the social network of Digg is a contagion process that generates many cascades. The submitter is the \emph{seed} of a cascade. However, there can be other means through which the story can reach a user. For instance, the user could independently find it on one of Digg's web pages or through a link from an external site. If a user find the story through other means than the {friend's interface}, she becomes an {independent seed} for another cascade. Not all seeds, however, generate non-trivial cascades. If a voter is unconnected or does not influence at least one of her fans to vote, the story does not spread. An independent user who generates a non-trivial cascade is its \emph{active seed}.

We used Digg API to collect data about 3,553 stories promoted to the front page in June 2009. The data associated with each story contains its title, id, link, submitter's name, submission time, list of voters and the time of each vote, and the time the story was promoted to the front page. In addition, we collected the list of voters' friends.\footnote{This data is available for research purposes at http://www.isi.edu/$\sim$lerman/downloads/digg2009.html.} We define an \emph{active user} as a person who votes for at least one story. In our data set there are 139,410 active users. Next, we get the connections between the active users. We say user $a$ is $connected$ to user $b$ if she is either a \emph{friend} or \emph{fan} of user $b$.  We store an active user in \emph{active users network} if she is connected to one or more active users. Hence, we are able to determine, whether a user who votes for a particular story is a fan of any previous voter.  Out of all active users, 69,524 users $connected$ to one or more active users. These 69,524 $connected$ users form the underlying \emph{friendship network}. Of these, 57,908 users form a giant connected component. $572$ of the $587$  distinct submitters belong to this friendship network.

We treat each story as an independent contagion process. We arrange all voters in the temporal order in which they voted for the story and extract the underlying social network of these voters.
Let $n_s$ be the number of \emph{active} seeds of the contagion process of a story $s$. We take each active seed to be independent of other seeds. Therefore, we can quantitatively characterize the cascade by an $n_s \times 1$ vector $\mathbf{c}_j$ for every node $j$ participating in the contagion process. Transmission rate $\alpha$ can be derived empirically from the network. In this work, without loss of generality, we set the value of $\alpha$ to 0.5.
We use the framework described above to study the macroscopic properties of cascades on Digg, such as the distribution of cascade size, diameter, etc. We also study the dynamics of evolution of cascades associated with some sample stories.

\subsection{Macroscopic Cascade Characteristics}
\label{sec:macro}
The stories in our data set generated 216,088 distinct information cascades on the Digg social network. Using the formalism described above, we calculate global properties of these cascades and plot their distribution. These properties include cascade size, spread, diameter, etc. Due to lack of space, we have included in this paper  just  some examples of the many properties that we can calculate using $\phi$.

To fit continuous distributions to discrete data, we treat a discrete distribution as if it was generated from a continuous probability density function and then rounded to the nearest integer. We do not use the commonly used methods such as least square minimization, because the data that spans many orders of magnitude and  least square minimization can produce substantially inaccurate estimation of parameters  of heavy-tailed distributions like the power-law~\cite{Clauset09}. We use Maximum Likelihood Parameter Estimation (MLE) to estimate the values of parameters for these distributions and KS statistics to test the goodness of fit. The closer the KS-statistics to 0, better the fit.  We study the following distributions: lognormal $F(x;\mu, \sigma)=0.5 \mbox{erfc}[-\frac{lnx - \mu}{\sigma \sqrt{2}}]$, Weibull $F(x;k,\lambda, \eta)=(1-e^{-{(\frac{x-\eta}{\lambda})}^k})$ , mixed Weibull   
$$F(x;\alpha_i, k_i,\lambda_i)=\sum_{i=1}^{n}\alpha_i(1-e^{-{(\frac{x}{\lambda_i})}^k_i})$$
with $\sum_{i=1}^{n}\alpha_i=1$, and power-law $$F(x;x_{min}, \alpha)= {(\frac{x}{x_{min}})}^{-\alpha+1}.$$
More often power law applies only for values greater than a certain minimum $x_{min}$. In such cases the \emph{tail} of the distribution follows the power law. Using the MLE estimates of $x_{min}$ and scaling parameter $\alpha$, we find what percent of the data comprises this {tail} of the distribution. We also investigated distribution fitting using the Double Pareto Lognormal distribution \cite{Reed03}  
\begin{eqnarray*}
F&(&x;\alpha, \beta,\mu, \sigma) = 0.5 \mbox{erfc}[-\frac{lnx - \mu}{\sigma \sqrt{2}}]  \\ 
 & - & \frac{1}{\alpha+\beta}(0.5 \beta x^{-\alpha} A(\alpha,\mu,\sigma) \mbox{erfc}[-\frac{lnx - \mu-\alpha \sigma^2}{\sigma \sqrt{2}}] \\
 &+& \alpha x^{-\beta} A(-\beta,\mu,\sigma)(1-0.5\mbox{erfc}[-\frac{lnx - \mu+\beta \sigma^2}{\sigma \sqrt{2}}])  )
\end{eqnarray*}
\noindent where $A(\theta,\mu, \sigma)=e^{\theta \mu+\frac{\theta^2\sigma^2}{2}}$.
Double Pareto Lognormal (DPLN) distribution  with $\alpha=2.8$, $\beta=1.9$, $\mu=3.0941$, $\sigma=0.3119$  gives the best fit for the number of cascades (better than any of the distributions  shown in Table \ref{tbl:mi} ) with likelihood of $-15.234 \times 10^3$ and $KS$ statistic of 0.0109.

\begin{figure} [htbp]
\scalebox{0.95}
{
\begin{tabular}{@{}c@{}c@{}}
  \includegraphics[height=1.5in]{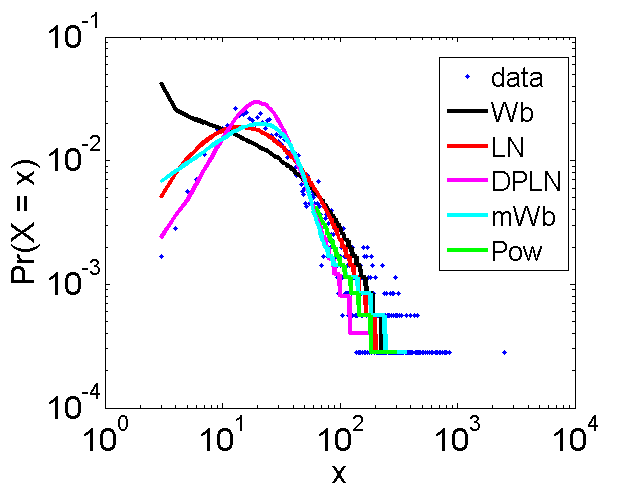} &
   \includegraphics[height=1.5in]{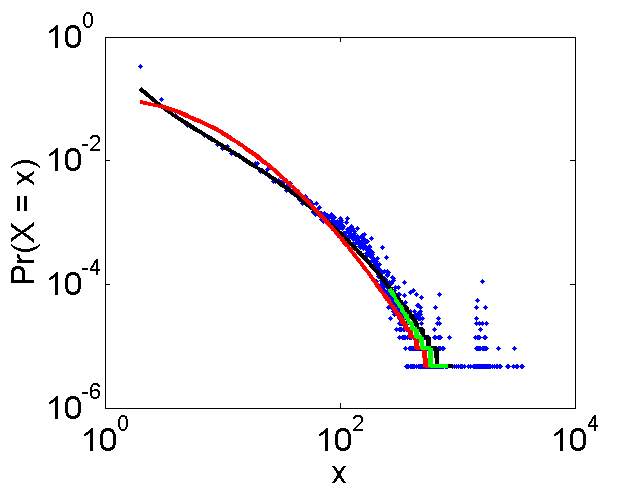} \\
  {num cascades} &  {cascade size} \\
  \includegraphics[height=1.5in]{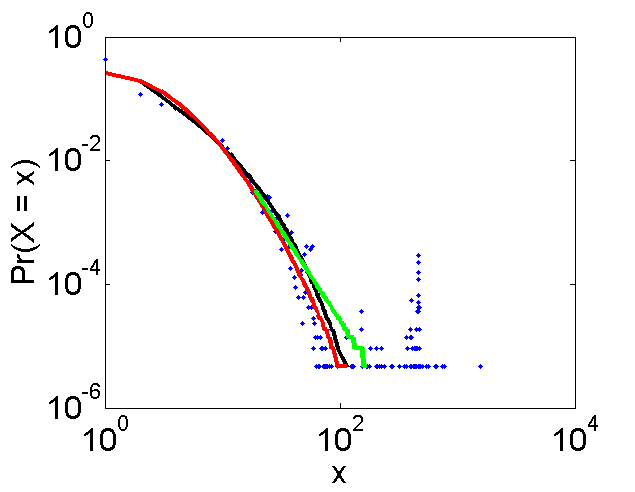} &
  \includegraphics[height=1.5in]{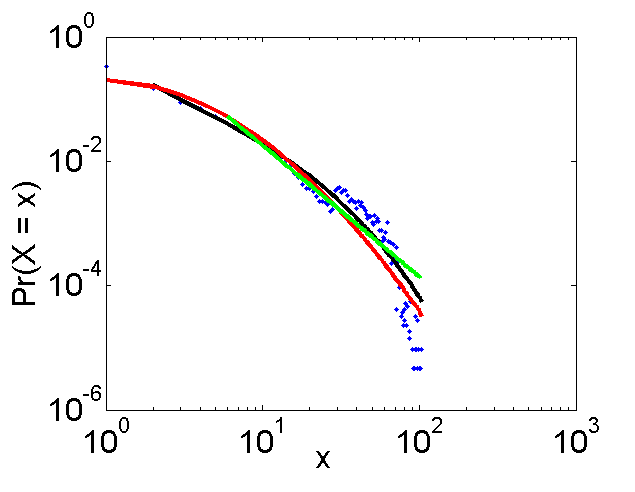} \\
{spread } &   {diameter } \\
  \includegraphics[height=1.5in]{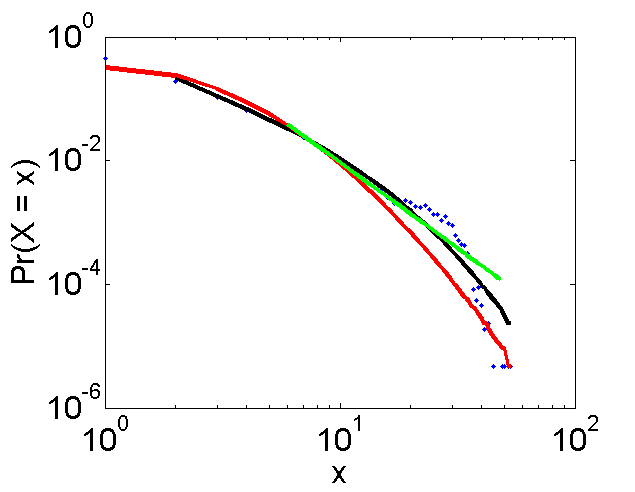} & 
  \includegraphics[height=1.5in]{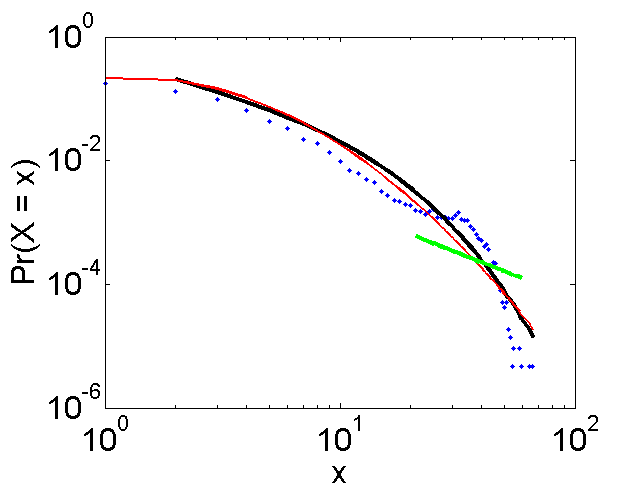} \\
{ave. path} & {log(num paths)}
\end{tabular}
}
\caption{PDF of distribution of cascade properties: number of cascades per story, cascade size, spread, diameter, average path length, and log of the number of paths. Distributions are fitted with the stretched exponential/Weibull (black), mixture of Weibull (cyan), lognormal (red)  and power law (green) functions. The double pareto lognormal distribution(magenta) gives a very good fit for the number of cascades.  }
\label{fig:cascade_metrics}
\end{figure}

Fig.~\ref{fig:cascade_metrics} shows the distribution of several macroscopic properties of the information cascades on Digg, along with functions that best describe them.  Table \ref{tbl:mi} shows the MLE estimates of these distributions.

\begin{table*}[htdp]
\caption{Parameter estimates for distributions that best describe data.}
\begin{center}
\scalebox{0.85}
{
\begin{tabular}{|c||c|c|c|c||c|c|c|c|c||c|c|c|c|c|}
\hline
&
\multicolumn{4}{c||}{\emph{Lognormal}} &
\multicolumn{5}{|c||}{\emph{Weibull (n=1)}} &
\multicolumn{5}{|c|}{\emph{Power Law}} \\
 \hline
	&	$\hat{\mu}$	&	$\hat{\sigma}$ 	&	$lk({10}^3)$	&	$KS$	&	$\hat{k}$	&	$\hat{\lambda}$	&	$\hat{\eta}$	 &	$lk({10}^3)$	&	$KS$	&	$\%$	&	$\hat{\alpha}$	&	$\hat{x_{min}}$	&	$lk({10}^3)$	&	$KS$		\\
	\hline
\# cascades	&	3.57	&	0.96	&	-17.57703	&	0.063	&	0.88	&	53.46	&	2.98	&	-17.90584	&	 0.1053
&	48.97	&	2.17	&	33	&	-9.01	&	0.291	\\
\hline																		
cascade size	&	2.06	&	1.43	&	-829.58593	&	0.175	&	0.41	&	7.44	&	1.24	&	-672.5123	&	 0.444
&	4.56	&	3.14	&	133	&	-55.1	&	0.036	\\
\hline																		
spread	&	0.94	&	1.00	&	-509.83819	&	0.255	&	0.59	&	2.47	&	0.83	&	-447.31692	&	0.56
&	12.21	&	2.92	&	10	&	-82.4	&	0.081	\\
\hline																		
diameter	&	1.19	&	1.14	&	-590.44924	&	0.186	&	0.55	&	3.43	&	0.91	&	-513.93646	&	 0.495
&	30.96	&	2.11	&	6	&	-234	&	0.690	\\
\hline																		
ave. path length	&	0.75	&	0.84	&	-431.7686	&	0.262	&	0.6	&	1.54	&	0.90	&	-342.79237	&	 0.79
&	15.01	&	2.78	&	5.88	&	-81.57	&	0.850\\
	\hline																		
log \# of paths	&	1.086	&	0.91	&	-349.58	&	0.392	&	0.717	&	3.107	&	0.848	&	-326.58	&	 0.673
&	2.51	&	1.5	&	21	&	-0.636	&	0.646\\
	\hline
\end{tabular}
}
\end{center}
\label{tbl:mi}
\end{table*}

We observe that lognormal or stretched exponential gives a good fit with the observed distribution, and that power law mostly (if at all) accounts for a small percentage at the tail of the distribution.
This indicates that a small number of core users may not be driving information propagation in online social networks on the whole. However, as the cascade size increases, some users may have disproportionate influence on information propagation.  Lognormal distribution indicates that the distribution might be generated by a multiplicative effect of many i.i.d random variables. Following the Fisher-Tippet-Gnedenko theorem, the stretched exponential distribution is the limit distribution of properly normalized extrema of a sequence of i.i.d random variables. Hence the distribution may have been generated by the extreme value of the a set of i.i.d random variables. A very good fit of the distribution of number of cascades with the DPLN distribution suggests a possible relationship between the distribution of number of cascades and geometric Brownian motion.  Future work includes, delving deeper into the probable causes of these distributions.

\subsection{Microscopic Cascade Characteristics}
The cascade generating function $\phi$ is an effective tool not only for computing the global properties of cascades, but  also for analyzing their microscopic dynamic signatures. In previous works, this was done by visualizing individual cascades~\cite{Leskovec07} or by creating a generative model of the contagion process~\cite{Leskovec07,Watts02}.
 Visualization, however, quickly becomes difficult, even for moderately-sized cascades. Generative models are \emph{ad hoc} in nature, and while they are designed to produce cascades with similar macroscopic properties as the observed cascades, they are not guaranteed to reproduce their microscopic characteristics. The cascade generating function, on the other hand, allows us to study microscopic properties of even very large cascades without the need to visualize them.

\begin{figure*}[tbh]
\begin{tabular}{@{}c@{}c@{}c@{}c@{}}
 \includegraphics[width=1.75in]{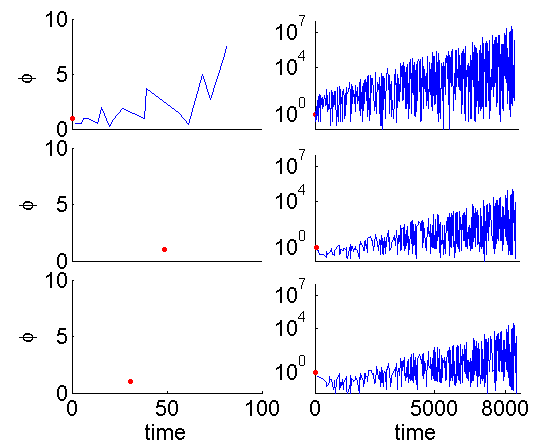} &
 \includegraphics[width=1.75in]{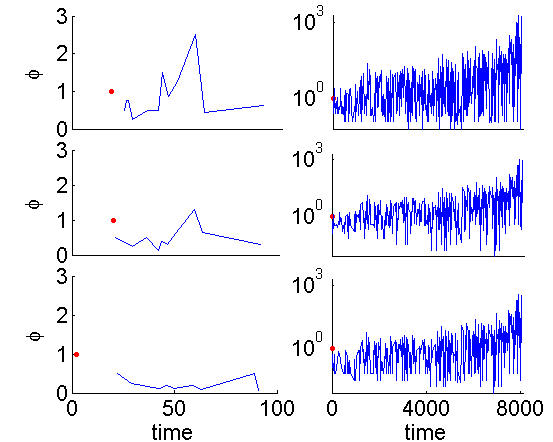} &
 \includegraphics[width=1.75in]{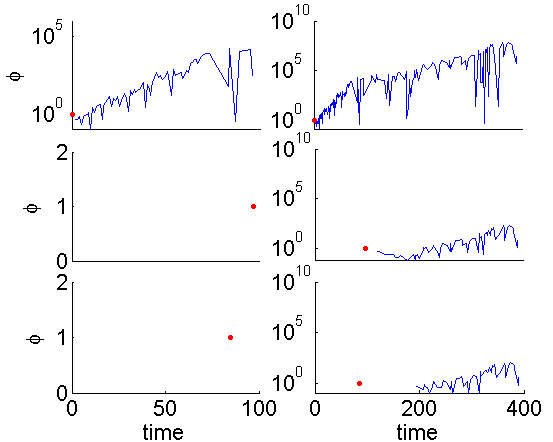} &
 \includegraphics[width=1.75in]{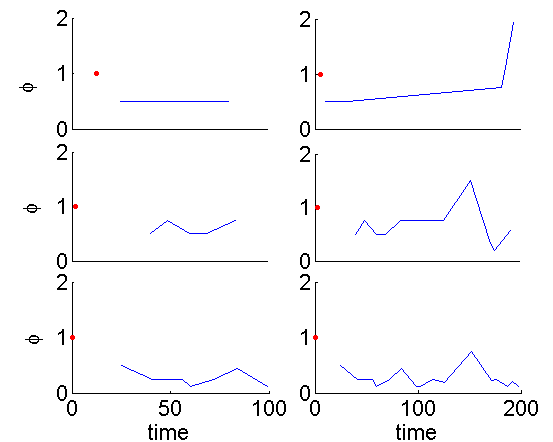} \\
 (a) Story 1 & (b) Story 2 & (c) Story 3 & (d) Story 4
 \end{tabular}
  \caption{Shows the cascade plot for  top 3 cascades for four stories. The left set of plots in each figure shows cascade evolution in the early stages of the contagion process, while the right set of plots shows cascade evolution over the entire time period. Red dot shows the time when cascade seed was activated.}
  \label{fig:phi}
\end{figure*}

We illustrate the use of cascade plots to study microscopic dynamics of cascades with four different stories. Story 1, titled ``Infomercial King' Billy Mays Dead at 50'' was submitted by a user who had 760 fans. This  story was among the most popular in our data set, receiving 8,471 votes, of which 1,244  were from fans. The contagion process of this story generated 853 cascades. Its diameter was 46, spread 412, and the average path length 24.
Fig.~\ref{fig:phi}(a) shows evolution of the cascade function $\phi(t)$ of the top three cascades, ranked by their largest $\phi$ value. The left-hand set of plots shows the early dynamics of the cascade ($t<100$), while the right-hand set of plots shows cascade dynamics over the entire time period. The seed of the cascade is shown in red.

The top cascade attains its largest value of $\phi=3.554\times 10^7$.
This cascade started early in the contagion process.  Though the seeds of the next two cascades were also activated within the first 100 votes, these cascades did not start growing until later.
Values of $\phi > 1$ imply that the voter is a fan of two or more previous voters. Large values of $\phi$ in Fig.~\ref{fig:phi}(a) indicate a \emph{community effect} (\emph{cf} Fig.~\ref{cascade_fig}(3)). This implies that information is spreading within an interconnected fan network.
Though initially the three cascades of Story 1 are very different, in their later stages, they become increasingly similar. This is due to mixing caused by ``{collision of cascades},'' which happens when the same nodes participate in different cascades.

The popularity of Story 2, titled ``Bender's back,'' is comparable to popularity of Story 1. Story 2  received 8,034 votes of which 1,464 were from fans and generated 722 cascades. Its  diameter was 26, spread 401, and the average path length 12. Fig.~\ref{fig:phi}(b) shows both the early and late-stage dynamics of the top three cascades generated by this story. However, the largest value of $\phi$ attained by any cascade was just $\phi=1859.4$, four orders of magnitude smaller than for Story 1. This indicates a much lower connectivity of the underlying fan network. In the three dominant cascades of this story, $\phi$ does not rise above 2.5 during the first 100 votes.
Low values of $\phi$ in the initial stages of cascade evolution imply a \emph{chaining effect}  (\emph{cf} Fig.~\ref{cascade_fig}(1)), or cascade growth by deepening. Unlike Story 1, here the seed of the dominant cascade is the $19^{th}$ voter.  However, as seen from the larger values of $\phi$ in the cascade plots, in the  later stages of information spread, community effect also comes into picture.

The third story in Fig.~\ref{fig:phi}(c) is titled ``Play Doctor On Yourself: 16 Things To Do Between Checkups.''
While this story was submitted by a  well-connected user (with 1,701 fans)  it did not become popular,
receiving only 390 votes of which 158 were from fans. This story generated 11 cascades, and its diameter was 48, spread 5, and the average path length 25. All of the first 100 voters participated in the dominant cascade, one initiated by the submitter himself. The maximum $\phi$ value reached by this cascade was very high ($\phi=7.53 \times 10^7$), even though this cascade was of short duration.   Unlike in  previous stories, we observed very high values of $\phi$ already within the first 100 votes, which indicates strong {community effect}, and high connectivity within the fan network.

For the final illustration we consider the story titled ``APOD: 2009 July 1 - Three Galaxies in Draco,'' shown in Fig.~\ref{fig:phi}(d).
The submitter of this story has only 27 fans. This story is one of the least popular in our data set, receiving only 199 votes, of which 27 were from fans. This contagion process generated eight cascades, its  diameter was 7, spread 7, and the average path length 2.6. In the early stages, constant values of $\phi$ in the dominant cascade (top plot in Fig.~\ref{fig:phi}(d)) indicate a \emph{branching effect} (\emph{cf} Fig. \ref{cascade_fig}(1)). This implies that cascade is growing in a star-like fashion, rather than deepening. The decreasing values of $\phi$ of the third cascade (bottom plot in Fig.~\ref{fig:phi}(d)) indicate a \emph{chaining effect}, implying that this cascade is deepening. We do not observe the \emph{community effect} either in the initial or later stages of this contagion process.
The maximum value of $\phi$ for this story is two and the average path length is 2.6, indicating that most of the voters are the fans of the submitter or submitter's fans, but are not themselves interconnected.

In summary, cascade plots can tell us much about the microscopic evolution of information cascade. Popular stories that have large participation also generated many cascades and had high spread. Initially they showed chaining and branching effects, as evidenced by $\phi$ values that are decreasing or staying constant in time, respectively. The community effect, manifested by growing values of $\phi$, is visible in later stages when a story penetrates and then spreads through a community. The trends of the dominant cascades grow increasingly similar with time due to the {mixing effect} of ``colliding cascades.'' However stories that do not become popular generate very few cascades and have low spread. When submitter is well connected, the community effect is visible in all stages of the contagion process, implying that the story spreads within submitter's community only. However, when submitter is poorly connected, cascades grow by chaining and branching.

\section{Related work}
Most of the earlier work does not clearly distinguish between cascades and the contagion processes generating these cascades. We believe that ours is the first work studying large scale cascades without link ambiguity. Though large scale studies of information cascades have been carried out earlier \cite{Leskovec05}, the cascades in general were small in size (O(10)). We on the other hand, have very large cascades (extending up to $O(10^4)$). The quantitative framework for analyzing cascades that we present here is very scalable and can be easily used to provide a efficient compressed representation of large cascades, when storing the complete information of the entire cascade is no longer trivial. In  previous studies \cite{Leskovec07, Leskovec05}, the authors characterize the cascades/contagion processes using a multilevel approach comprising of  global  and local signatures. Cascades are considered to be approximately isomorphic if they have the same  global signature. If the global signatures match, more expensive isomorphism tests based on local signatures are carried out. To aid reasoning about cascades, the authors focus on \emph{local cascades}, which they define as the `cascade in the (undirected) \emph{neighborhood} of the node', which for every node is the subgraph induced on the nodes reachable from it. They enumerate the shapes of these local cascades. As cascades grow in size, the number of possible shapes increases exponentially and such enumeration becomes infeasible. Note that in this work we provide a scalable, efficient  and compressed representation of the observed cascades and make no claims about whether the observed cascades are the actual cascades or fragments of them.

In our study of Digg,  we have cascades of size up to $\approx 20,000$. We aim to deduce their qualitative as well as quantitative properties, such as shape and size.   Hence a more formal framework for characterizing cascades is required. In this paper we provide such a formalism, which not only captures the macro and micro level signatures described above, but much more.  For instance, it captures the similarity between cascades, which were initially similar but later become dissimilar; or, similarity between cascades that are similar in some stage of their growth. This formalism enables us to distinguish between cascades and obviates the need for enumerating them or drawing their shape.

In \cite{ Rodriguez10, Gruhl04}, the underlying network on which information spreads is not observed, but has to be inferred from the observed cascades. However such inferences \cite{Rodriguez10} are based on the hypothesis that the contagion process follows an independent cascade model~\cite{Kempe03}.
Our work, on the other hand, focuses on providing a quantitative tool to analyze the  trends and patterns of actual contagion processes observed on real-life networks. Even when the underlying network is predicted using a different inference methods, e.g.,~\cite{Rodriguez10,Gruhl04}, the trends of the contagion process occurring on the network can be investigated using the cascade generating function. Future work includes using these tools to aid the verification  or  rejection of the hypothesis used for modeling information spread~\cite{Bailey75, Dodds04, Watts02}.  It can also prove to be an effective tool to evaluate the robustness of inferred networks~\cite{Rodriguez10}.

As demonstrated by the third story in our examples, we observe that if the submitter is well connected, the community effect is visible at all stages. However, initial popularity  only within the tightly knit community  (shown by a high cascade value  and few seeds in the initial stages) does not ensure global popularity (large number of votes). In contrast, stories submitted by a not so well connected user, which spreads by branching and deepening initially (with low cascade values), but  have larger number of initial active seeds  become more popular globally (as shown by the second story in the example). This observation is in agreement to those reported in \cite{Lerman08wosn,Bakshy09} that content diffusing primarily through an interconnected community tends to be confined to that community.
These cascades are also complicated by the interplay between social influence and homophily~\cite{Anagnostopoulos08,Choudhury10}. Future work will address these questions more closely.

In previous works~\cite{Leskovec07,Leskovec05}, the cascade size was found to be described well by the power-law distribution. However, we observe that power-law only accounts for a small fraction of cascades at the tail of the distribution. Rather, the entire data can be approximated well with a stretched-exponential (weibull), lognormal or double pareto lognormal distributions, similar to those observed in \cite{Guo09}.

\section{Conclusion}
In this paper we adopt an empirical approach to study cascades on networks.
We believe that our work is first to provide a mathematical framework to quantify and analyze cascades, even for applications requiring real-time or online analysis. The mathematical framework is based on the cascade generating function, which quantitatively characterizes both the micro and the macroscopic properties of the cascade. The macroscopic properties that can be efficiently calculated using this tool include the diameter and the spread of the cascades. This function also  provides an efficient compression of the information encoded in cascades. In spite of having pseudo-linear space complexity, it can be used to reconstruct the shape of the cascade with high degree of accuracy.

Although large scale studies of cascades have been carried out, the size of cascades in these studies was relatively small. To the best of our knowledge, this is the first study of very large cascades with thousands of participants.
We use this function to study information cascades on an online-social news aggregator Digg. For macroscopic properties like number of cascades in a contagion process, cascade size, spread, diameter, average length and so on, we observe a stretched exponential (Weibull) \remove{mixture of weibull} or a lognormal distribution fits well with the observed distribution. Double Pareto Lognormal distribution gives a very good fit for the distribution of number of cascades. Usually power law accounts (if at all) for a small percentage of data in the tail of the distribution. Microscopic analysis also revealed interesting insight to cascades and contagion processes, such as the possible effect of the initial number of seeds and of the branching, chaining and community effect on the initial popularity of news.

\subsection*{Acknowledgments}
This work is supported in part by the Air Force Office of Scientific Research and in part by the National Science Foundation under award 0915678.
{

\comment
{
\appendix{}

\comment
{
\begin{algorithm}
\caption{ Efficient computation of Diameter}
\label{alg:4}
\begin{algorithmic}
\small
{
\STATE{\bf{Input}}\\
$A$: Adjacency Matrix of the Cascade Graph   \\
$N$: number of nodes in A
\STATE{\bf{Output}}\\
$B_{\alpha}$: Minimal Probability Matrix ($N \times k$),$D_{\alpha}$: Diameter Matrix ($N \times k$)  \\
$\forall p \in [1,k]$, let $j$ be the time step at which seed $p$ is activated.
\STATE{\bf{Initialize}}\\
$B_{\alpha}(i,p)=\infty  \forall i,p$, $D_{\alpha}(i,p)=\infty  \forall i,p$\\
 \FOR{$i= 2$ to $N$}
            \IF {$A_c(i,j)!=0$  and $i==j+1$}
                \STATE $B_{\alpha}(i,p)=\alpha A_c(i,j)$
                 \STATE $D_{\alpha}(i,p)=\alpha A_c(i,j)$
               \ENDIF

               \FOR{$\forall p$  and $i\ge j+2$}
              \IF {$A_c(i,j)!=0$}
                \STATE  $B_{\alpha}(i,p)=  \alpha A_c(i,j)$
                \STATE $D_{\alpha}(i,p)=\alpha A_c(i,j)$
               \ENDIF

              \FOR{$k = 1$ to $i-j-1$}
             \IF {$ A(i,i-k)!=0$ and  $ B_{\alpha}(i-k,p)!=0$ }
                \IF {$ B_{\alpha}(i,p)> \alpha A(i,i-k) B_{\alpha}(i-k,p) $ }
                        \STATE$B_{\alpha}(i,p)= {\alpha A(i,i-k) B_{\alpha}(i-k,p)} $
                      \STATE $D_{\alpha}(i,p)= \alpha A(i,i-k) (  B_{\alpha}(i-k,p)+ D_{\alpha}(i-k,p)) $
             \ENDIF
             \ENDIF
              \ENDFOR
              \ENDFOR
   \ENDFOR
   }
\end{algorithmic}
\end{algorithm}
}
}
}

\balancecolumns 
\end{document}